\documentclass[aps,pre,floats,twocolumn,twoside,floatfix,superscriptaddress]{revtex4}

\usepackage{graphicx}
\usepackage{amsmath,amssymb}
\usepackage{psfrag}
\usepackage[utf8]{inputenc} 

\newcommand{\Tc}{T_{\scriptscriptstyle \rm c}}
\newcommand{\Tp}{T_{\scriptscriptstyle \rm perc}}
\newcommand{\nuH}{\nu_{\scriptscriptstyle\rm H}}

\newcommand{\HFK}{H_{\scriptscriptstyle\rm p}}
\newcommand{\Hg}{H_{\scriptstyle\rm g}}
\newcommand{\tauFK}{\tau_{\scriptscriptstyle\rm p}}
\newcommand{\taup}{\tau_{\scriptscriptstyle\rm perc}}
\newcommand{\taug}{\tau_{\scriptscriptstyle\rm g}}
\begin{document}

\title{Domain size heterogeneity in the Ising model: geometrical and thermal
transitions}

\author{André R. de la Rocha}
\affiliation{Instituto de F\'\i sica, Universidade Federal do
Rio Grande do Sul, CP 15051, 91501-970 Porto Alegre RS, Brazil}

\author{Paulo Murilo C. de Oliveira}
\email{pmco@if.uff.br}
\affiliation{Instituto Mercosul de Estudos Avançados, Universidade Federal da
Integração Latino Americana, Foz do Iguaçu, PR, Brazil}
\affiliation{Instituto de Física, Universidade Federal Fluminense, Niterói, RJ, Brazil}

\author{Jeferson J. Arenzon}
\email{arenzon@if.ufrgs.br}
\affiliation{Instituto de F\'\i sica, Universidade Federal do
Rio Grande do Sul, CP 15051, 91501-970 Porto Alegre RS, Brazil}

\date{\today}

\begin{abstract}
A measure of cluster size heterogeneity ($H$), introduced 
by Lee et al [Phys. Rev. E {\bf 84}, 020101 (2011)] 
in the context of explosive percolation, was recently applied to random percolation
and to domains of parallel spins in the Ising and Potts models. 
It is defined as the average number of different domain sizes in a given 
configuration and a new exponent was introduced to explain its scaling 
with the size of the system. In thermal spin models, however,
physical clusters take into account the temperature-dependent correlation between
neighboring spins and encode the critical properties of the phase transition.
We here extend the measure of $H$ to these clusters and, moreover, present new 
results for the geometric domains for both $d=2$ and 3. We show
that the heterogeneity associated with geometric domains has a previously
unnoticed double
peak, thus being able to detect both the thermal and percolative transition. 
An alternative interpretation for the scaling of $H$  that does not 
introduce a new exponent is also proposed.
\end{abstract}

\maketitle
\section{Introduction}

Cluster or domain size distributions are commonly used in statistical mechanics
to unveil geometric properties and characterize both the equilibrium~\cite{CaZi03}
critical behavior and the out of equilibrium dynamics~\cite{ArBrCuSi07,SiArDiBrCuMaAlPi08} of several models.
If the system is finite, not all possible domain sizes are present on a 
single configuration. These sample-to-sample fluctuations disappear and 
the distributions become dense in 
the infinite size limit or after ensemble averages are taken. 
In order to characterize these fluctuations, a  quantity associated 
with how heterogeneously sized the equilibrium domains are, $H$, was recently 
proposed in the context of explosive percolation~\cite{LeKiPa11} and then 
applied to ordinary percolation~\cite{NoLePa11} and to the 
Ising~\cite{JoYiBaKi12} and Potts models~\cite{LvYaDe12}. Differently from
the cluster 
size distribution, the heterogeneity $H$ only takes into account whether a given 
size is present in each configuration and gives the number of such 
distinct sizes. Although
the number of equal-sized clusters and their actual size do not enter in the measure of $H$, 
both entropic and thermal effects are relevant. At high temperatures, thermal
noise breaks large clusters and the fragments are small, and so is the
diversity and $H$. On the other hand, for low temperature, the presence of a 
very large, perhaps percolating, cluster leaves small room to the other clusters, 
also decreasing the diversity.

We here revisit the original formulation for the equilibrium scaling behavior
of $H$ for the Ising model and present new results for 
both geometric and physical clusters~\cite{FoKa72,CoKl80}. While the former 
considers all nearest neighbor parallel spins belonging to the same domain, the latter only 
takes the fraction of spins effectively correlated. This difference, although important for 
systems of interacting spins under thermal noise, obviously does not exist for percolation 
models. For random percolation~\cite{NoLePa11}, $H$ has a clear peak and an excellent 
data collapse is obtained. For the Ising model,
no peak has been observed for the system sizes considered in Ref.~\cite{JoYiBaKi12}. 
Although in 2d both
kinds of spin domains percolate at the same temperature, only the physical ones 
encode the critical properties (exponents).
It is thus interesting to see how their heterogeneity, $\HFK$, differs from $\Hg$ 
measured with the geometric domains.  As we show below, even for small sizes,
$\HFK$ presents a clear peak and an excellent collapse is obtained. 
For the geometric domains, on the other hand, we observe two peaks for sufficiently
large systems, none of which were observed in Ref.~\cite{JoYiBaKi12}, where only 
an abrupt change in the derivative was present at the critical temperature $\Tc$. 
The first peak is small, very close to $\Tc$, while the second one is large, broad and appears
far above the critical temperature. Although the very existence of such double
peak is interesting enough, there are several questions that must be
answered. What is the origin of such peaks? Do they merge, leaving a
single peak, as the system size diverges? Do they share the same exponents or the percolation
exponents play a role? What happens in 3d where the critical temperature
does not coincide with the percolation threshold of geometric domains? 
In order to try to answer these questions, we studied the equilibrium Ising model on 
square and cubic lattices with linear sizes up to 2560 and 250, respectively. 
Averages up to 200 samples were performed for the smaller systems while larger sizes 
required fewer samples. All measures were obtained after equilibrating the system 
through 500 Swendsen-Wang steps~\cite{NeBa99}.

The paper is organized as follows. We start Sec.~\ref{section.H} by reviewing the 
definition of $H$ and its scaling behavior in the context of a thermal system. 
Then, in Secs.~\ref{section.2d} and \ref{section.3d} we present the results of
extensive simulations for the 2d and 3d Ising model, in 
which we measure $H$ for both geometric and physical clusters.  We finally 
discuss, in Sec.~\ref{section.conclusions}, our results and conclude.

\section{Size Heterogeneity}
\label{section.H}

Differently from the susceptibility that corresponds, in percolation terms, to the 
mean cluster size~\cite{StAh94}, and thus correlates with both the size and frequency of fluctuations, 
for the measure of $H$ it is only the presence of a given cluster size that is relevant. Both at 
too low and too high temperatures, the distribution of 
cluster sizes is concentrated at small domains, the latter because the large
noise dissolves the clusters and the former because a giant cluster
dominates the system leaving almost no room for the others. In both cases
the size distribution has a small tail, domains are rather homogeneous and $H$ 
is not large. We thus expect a maximum of $H$ between these two limits, close to 
the critical temperature because the fragmentation of the infinite cluster creates 
several small domains and the size distribution per site, $n(s)$, being
broad (power law), will increase the probability of these clusters having different
sizes and thus contribute to $H$.  
Moreover, as the linear size $L$ increases, more
and larger clusters are allowed when the average distribution is broad enough and $H$
must also increase with system size. Notice, however, that having a broad distribution
does not necessarily imply a large value of $H$: it depends on how many of these 
different sizes are present on a given state or, equivalently, the amount of
vacancies in the distribution for a particular configuration of a finite size
system (the average cluster size distribution, on the other hand, being
an ensemble average, is dense). 

\begin{figure*}[htb]
\includegraphics[width=5.9cm]{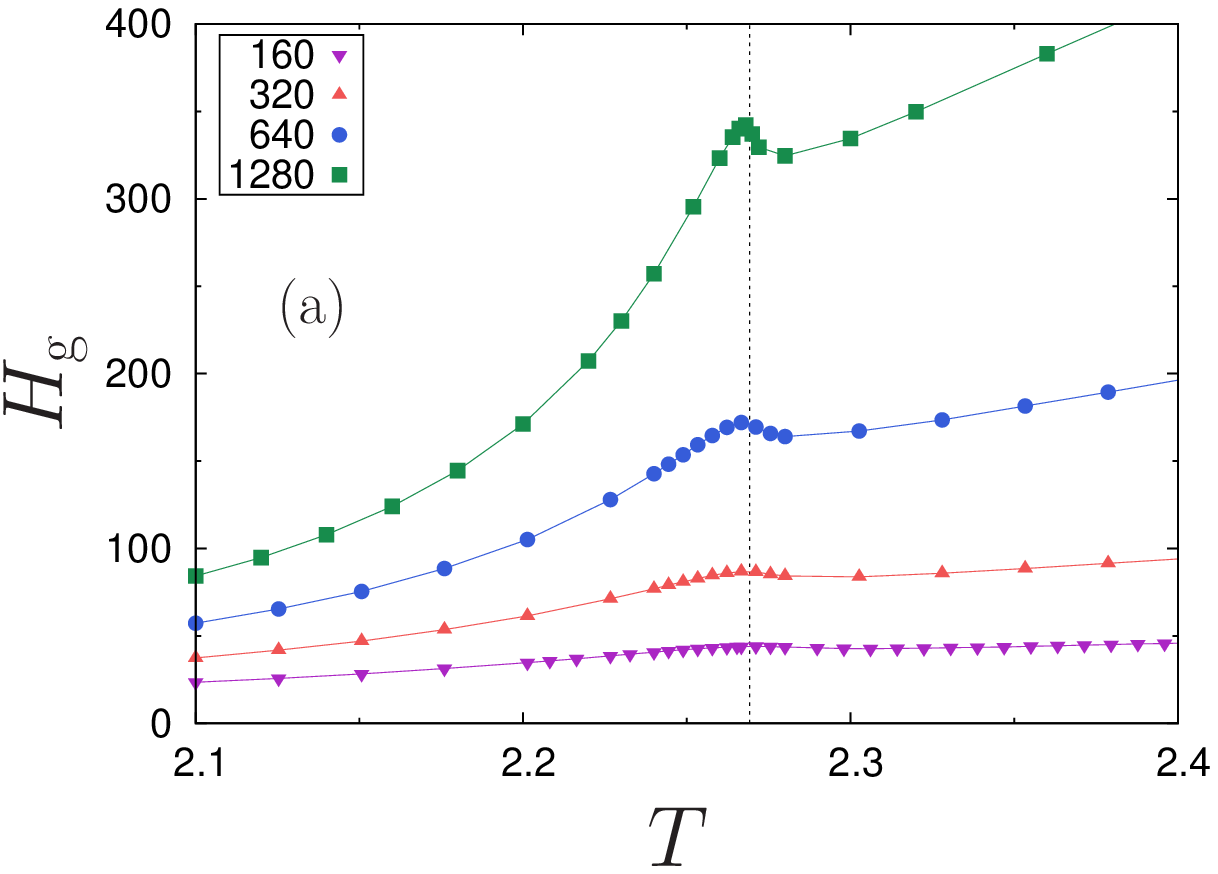}
\includegraphics[width=5.9cm]{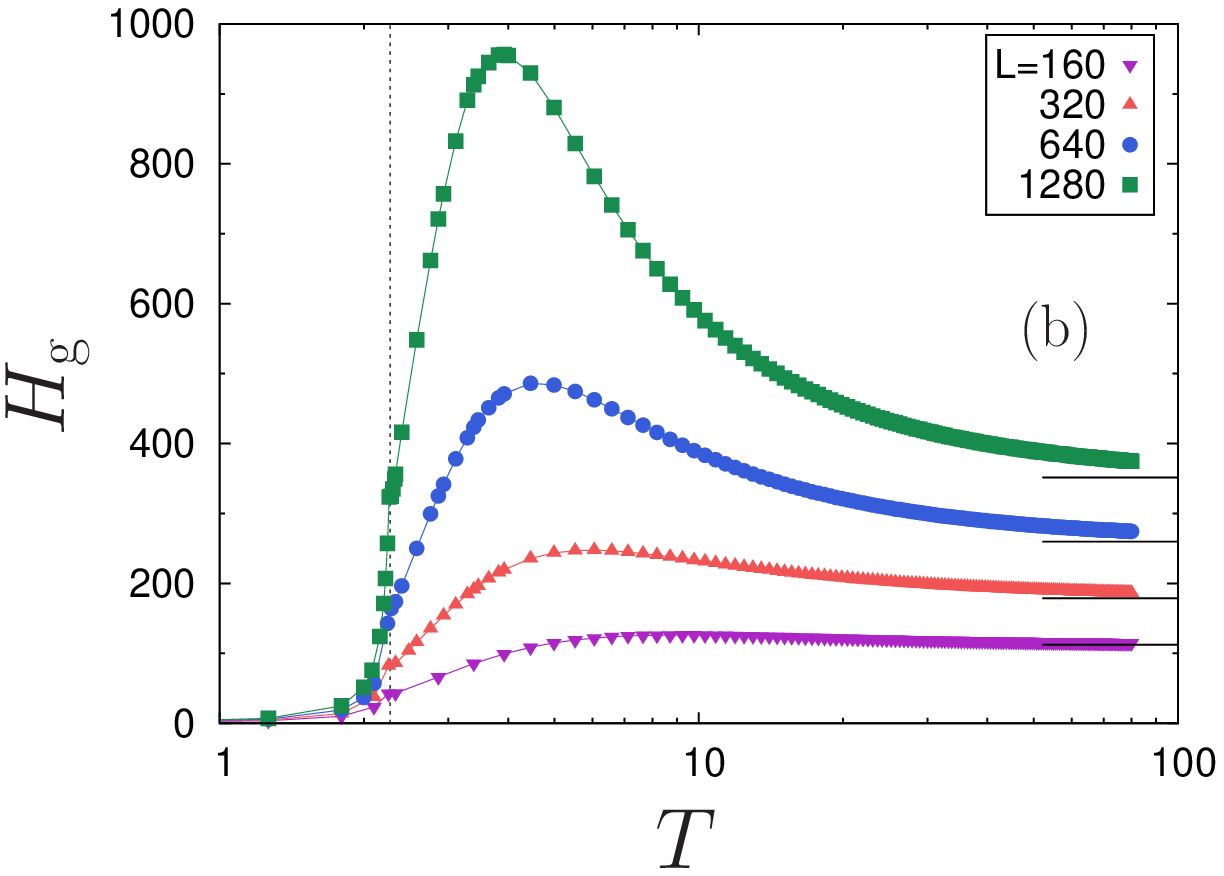}
\includegraphics[width=5.9cm]{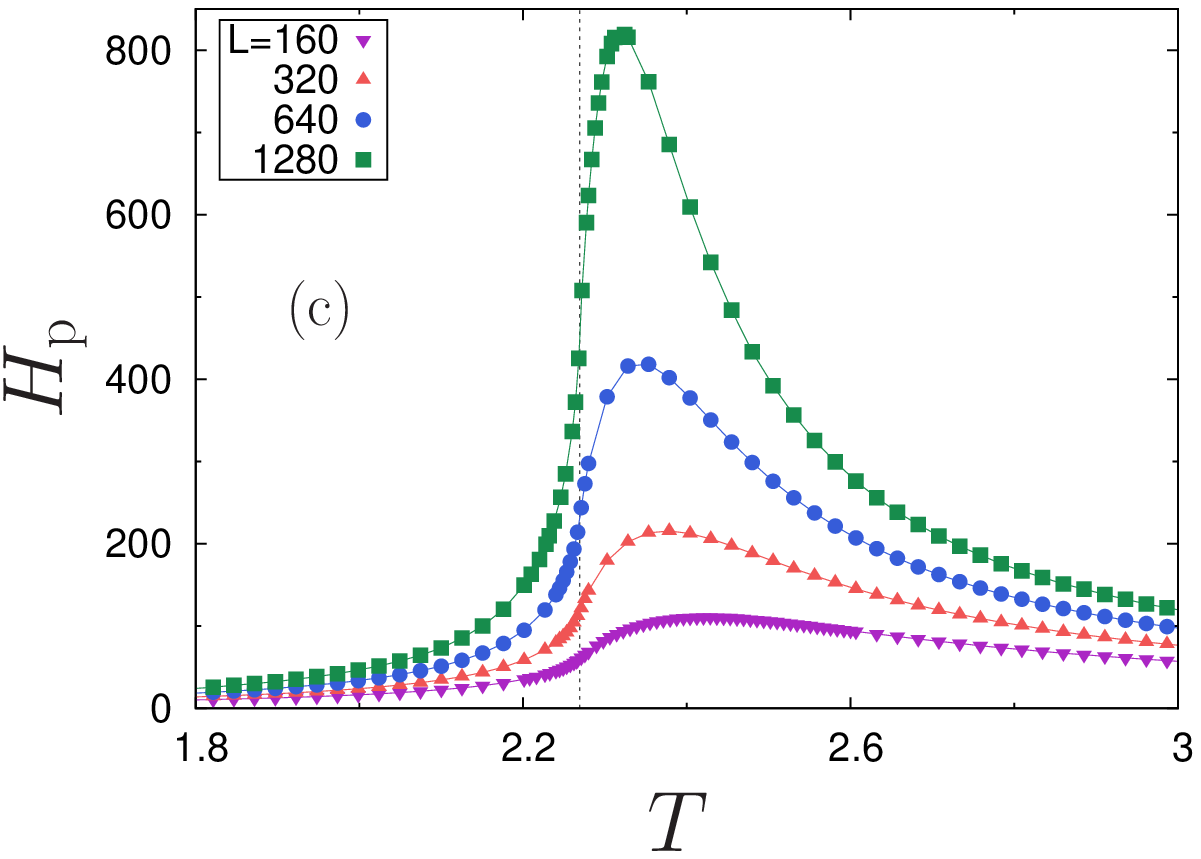}

\includegraphics[width=5.9cm]{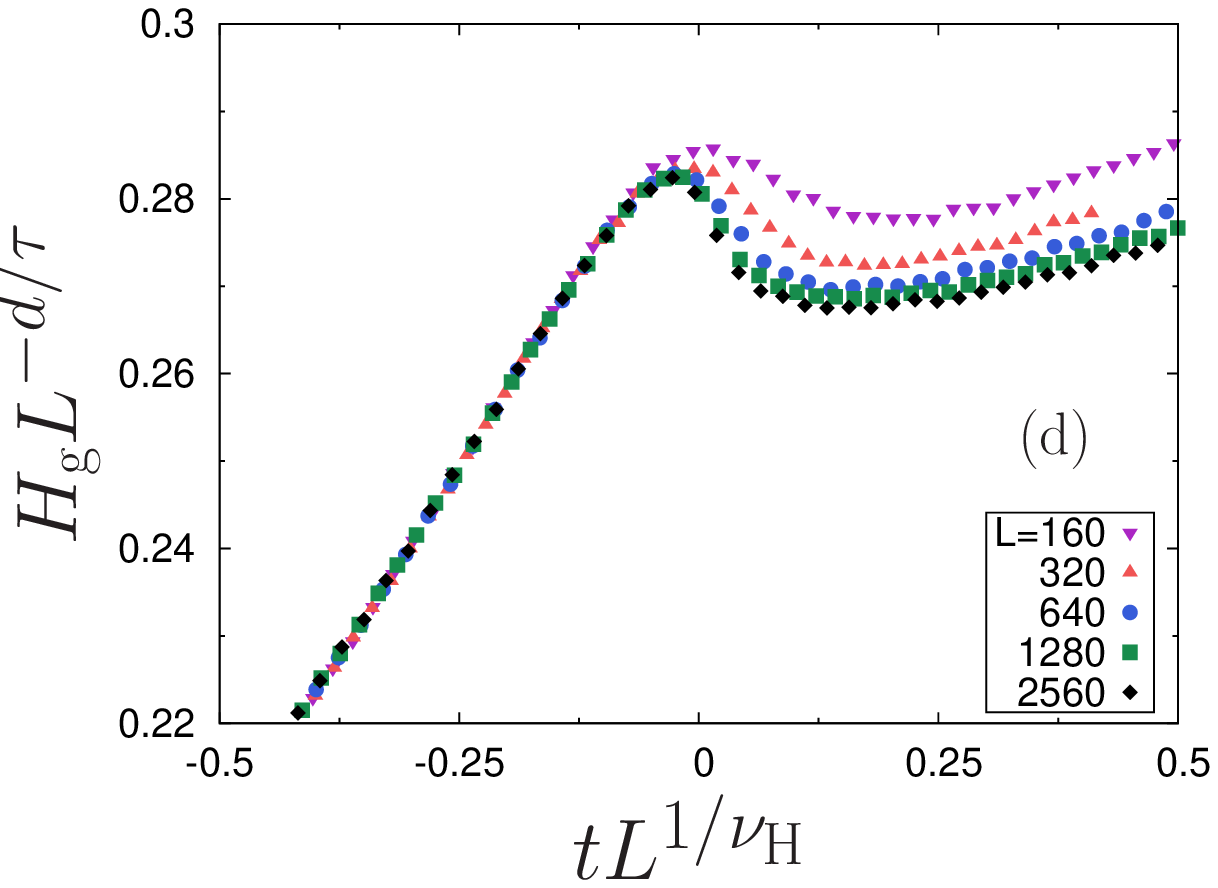}
\includegraphics[width=5.9cm]{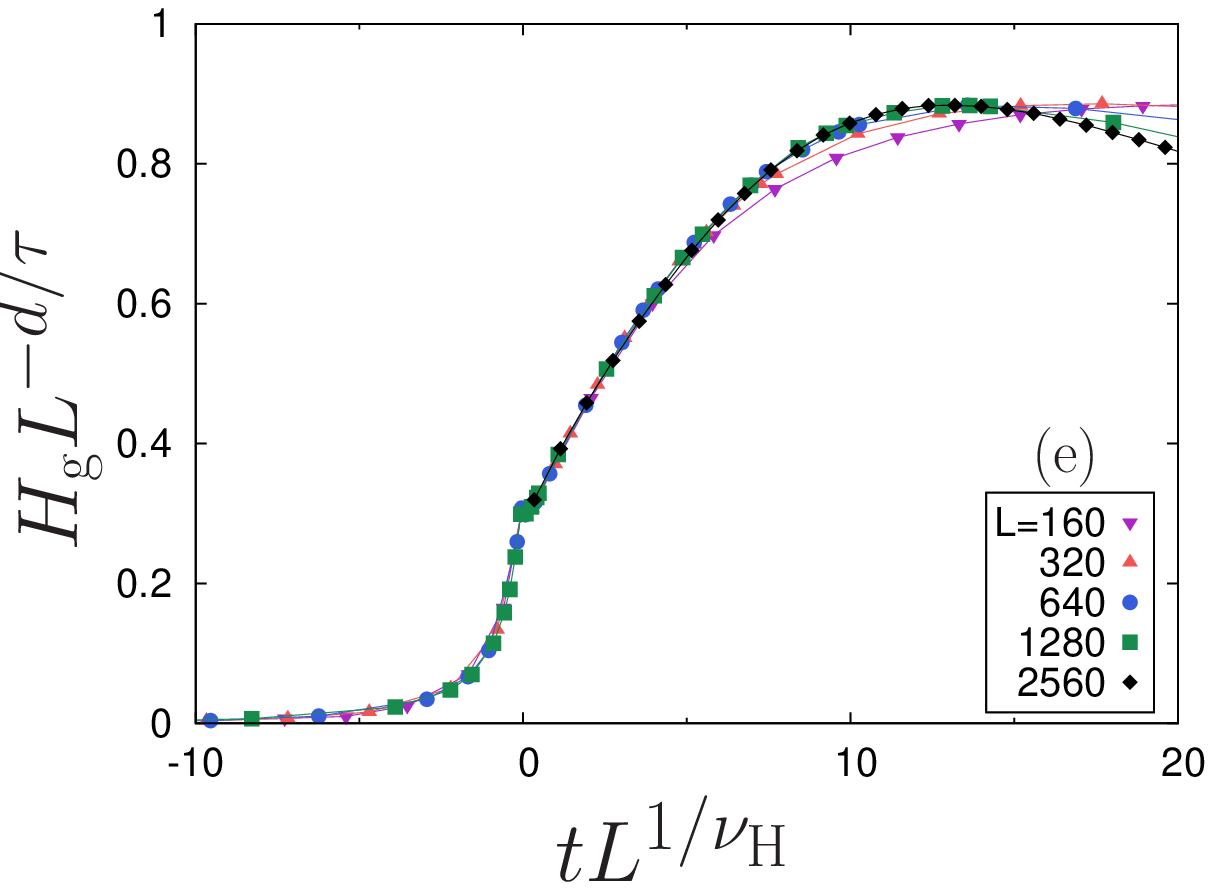}
\includegraphics[width=5.9cm]{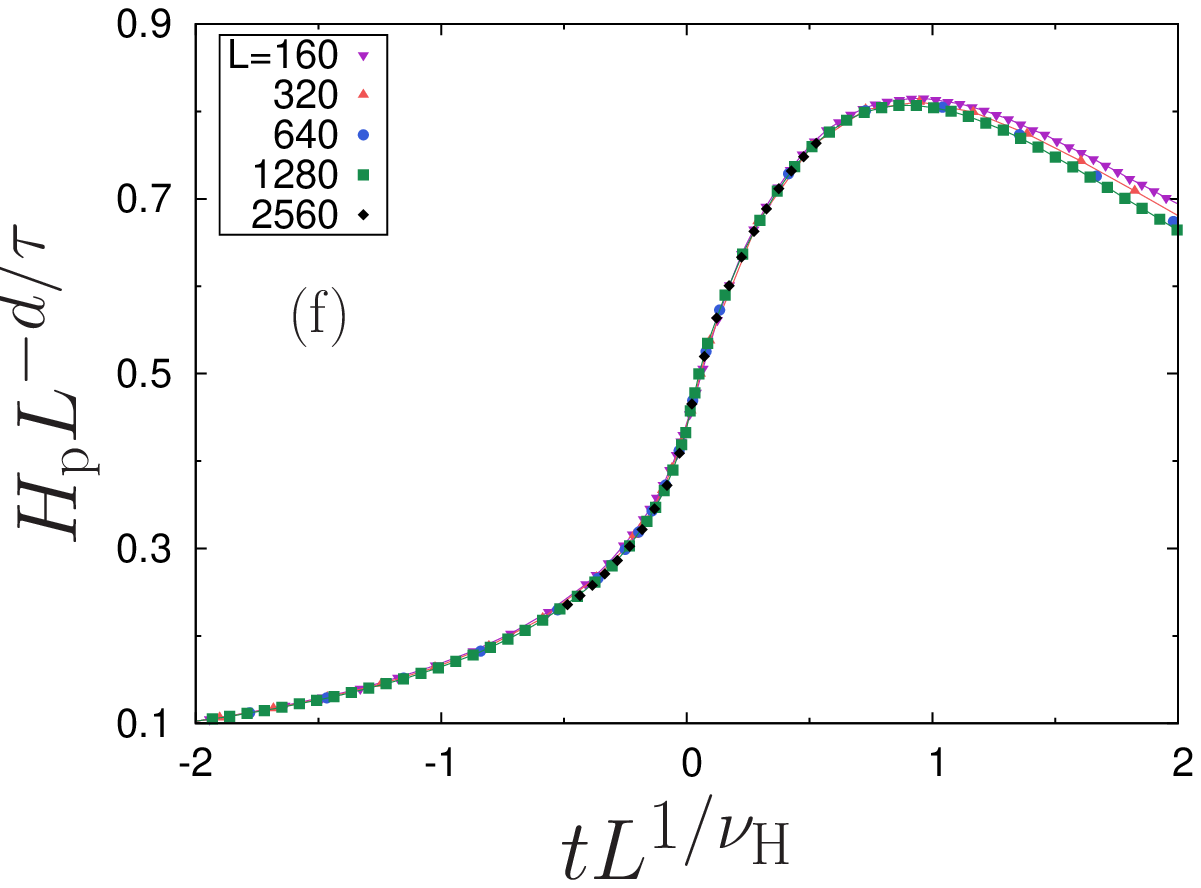}
\caption{Size heterogeneity $H$ for the $d=2$ Ising model 
in equilibrium as a function of the temperature for several system 
sizes (top) and the corresponding collapses (bottom) using
the reduced temperature $t\equiv T/\Tc-1$. 
The vertical dotted line locates the critical temperature $\Tc=2/\ln(1+\sqrt{2})$.
(Left) First (small) peak for the geometric domains, very close 
to $\Tc$. The best collapse, keeping $\nu=1$, was obtained with $\tau\simeq 2.016(4)$.
(Middle) Second, large and broad peak for the geometric domains. The height of
the peak grows with an effective $\tau\simeq 2.048$ that is definitively
different from the first peak. A good collapse is obtained with $\nu\simeq 1.25$. 
The horizontal lines correspond to an independent measure of $H$ at $T\to\infty$ using
a random configuration. 
(Right) The same but for the physical domains whose exponents are $\tauFK =31/15$
and $\nuH=31/16$. In this case, there is a single peak close to $\Tc$.
Notice the different horizontal scales in all graphs and the logarithmic one in (b).}
\label{fig.H.eq.Ising}
\end{figure*}

The precise dependence of $H$ on the linear size $L$ can be understood with
a scaling argument~\cite{LeKiPa11,NoLePa11}. Let us start with the height of the
peak close to the critical temperature. Small clusters are present with a high probability in 
a finite size system and we define $s_0$, which is a function of the
temperature, as the (average) smallest size not present in a configuration. Below $s_0$ the distribution is dense, 
that is, the expected number of clusters whose sizes are smaller than $s_0$, $L^d n(s<s_0)$, is larger than 
one. On the other hand, above $s_0$ there are some sizes without a realization, 
$L^d n(s>s_0)<1$ and vacancies appear in the distribution. Thus, by definition,
$L^d n(s_0) \sim 1$. As we will see, the contributions to $H$ from clusters in these two regions 
are of the same order~\cite{NoLePa11}. Close to the transition, we
write~\cite{StAh94} $n(s)\simeq s^{-\tau}f(z)$ 
where $z=ts^{\sigma}$ and $t\equiv T/\Tc-1$ is the reduced temperature. Two critical exponents 
are introduced. The first one, $\sigma$, is associated with the extension of the 
power-law tail as the criticality is approached while the
Fisher exponent $\tau$ is the exponent of such power-law. For 
the 2d cases considered here, we have
$\taug=379/187$ for geometric domains~\cite{StVa89,JaSc05a}, $\tauFK=31/15$
for physical clusters and $\taup=187/91$ for random percolation. The scaling function $f(z)$
approaches a constant for $|z|\ll 1$ and has a fast decay for $|z|\gg 1$.
Close to the transition, $n(s_0)\sim s_0^{-\tau}\sim L^{-d}$, and thus $s_0$ scales 
as $s_0\sim L^{d/\tau}$~\cite{LeKiPa11}. The remaining contribution to $H$
comes from the clusters with $s>s_0$. These clusters may be ranked 
by size, starting from the largest~\cite{JaStAh98,NoLePa11}. At the criticality,
the largest cluster size lies in the interval $[s_1,\infty)$ such that
the expected number of clusters in this interval is unity. That is, the
integral of $n(s)$ from $s_1$ to $\infty$, multiplied by the size of the system
is 1. Analogously, the second largest cluster lies in the interval $[s_2,\infty)$ 
such that the integral of $n(s)$ from $s_2$ to $\infty$, 
multiplied by the size of the system, is now 2. Repeating the process, for the 
$r$-th cluster in the rank, at the critical temperature, we get
\begin{equation}
r = L^d \int_{s_r}^\infty ds\; s^{-\tau} f(0) \sim L^d s_r^{1-\tau}.
\label{eq.ranking}
\end{equation}
In the appendix we present a more detailed discussion of the convergence
of this integral.
We get that $s_r \sim (L^d/r)^{1/(\tau-1)}$ and the fractal dimension of these large
clusters is $d/(\tau-1)$~\cite{StAh94}, larger than $d/\tau$ of $s_0$. 
As $r$ increases, remembering that we are ranking the sizes in reverse
order, $s_r\to s_0$. Since $s_0\sim L^{d/\tau}$, for very large $r_0$ one has 
$L^{d/\tau} \sim (L^d/r_0)^{1/(\tau-1)}$, hence $r_0 \sim L^{d/\tau}$.
Thus, the two contributions to $H$, $H\sim s_0+r_0$, scale in the same way 
and $H\sim L^{d/\tau}$~\cite{NoLePa11}. 
Since $H$ and $s_0$ share the same scaling behavior, measuring the first 
missing cluster size $s_0$ is enough to obtain the critical scaling behavior of $H$.

The point at which the distribution deviates from the power law grows as 
$s_*\sim |t|^{-1/\sigma}$, what defines the exponent $\sigma$. 
As pointed out in Ref.~\cite{JoYiBaKi12}, the competing scales in this 
problem can be considered $s_0$ and $s_*$ instead of the usual $L$ and $\xi$. 
When $s_*\gg s_0$, the power law is well developed and the distribution for
a single configuration has many holes (in the opposite limit, $s_*\ll s_0$,
such distribution is dense and the system behaves as if its size was infinite). 
In this case, $H\sim s_0\sim
L^{d/\tau}$. Away from the transition, $s_*$ decreases and the critical region 
extends up to the point in which $s_*\sim s_0$. Thus,
$L^d s_*^{-\tau}\sim 1$ and, using the above definition, the size of
the critical region is $|t|\sim L^{-d\sigma/\tau}\equiv L^{-1/\nuH}$,
where  $\nuH=\tau/\sigma d$ or, equivalently, $\nuH=\nu\tau/(\tau-1)$. 
Consistently with this,
Lee {\it et al}~\cite{LeKiPa11} considered 
that the scaling behavior of $H$ was given by $H(t,L)=L^{d/\tau} h(|t|L^{1/\nuH})$. 
While close to the transition the scaling function $h(x)$
is a constant, away from $\Tc$ the behavior is $h(x)\sim x^{-1/\sigma}$.

\subsection{2d}
\label{section.2d}

The definition of $H$ and its scaling were proposed and verified in Ref.~\cite{LeKiPa11} 
for explosive percolation and, later on, also in ordinary percolation~\cite{NoLePa11} and some
2d spin models~\cite{JoYiBaKi12,LvYaDe12} for which there are several possible
definitions of domains. For the Ising model, on which we focus here, a (geometric) 
domain is defined as a connected group of nearest neighbors aligned spins (we do not 
distinguish, as in Ref.~\cite{JoYiBaKi12}, between up and down spins clusters). 
We show, in Fig.~\ref{fig.H.eq.Ising}, the geometric domains heterogeneity
$\Hg$ as a function of the temperature for several system sizes near 
the 2d Ising critical point. For the small sizes considered in Ref.~\cite{JoYiBaKi12},
only a change in the derivative of $\Hg$, very close to $\Tc$, was observed. 
By considering larger sizes, interestingly, two well separated peaks develop. 
The first, small one (Fig.~\ref{fig.H.eq.Ising}a), is very close to $\Tc$
and, as shown by the collapse in Fig.~\ref{fig.H.eq.Ising}d, grows as $L^{d/\tau}$. 
A good collapse was obtained with $\tau\simeq 2.016(4)$ (keeping $\nu=1$)
and, accordingly,  $\nuH\simeq 1.984$. These values are close to the
exponents for the geometric domains in the Ising model~\cite{JoYiBaKi12},  
$\taug=379/187\simeq 2.027$ and $\nuH=379/192\simeq 1.974$. 
On a broader range of temperatures (notice the logarithmic scale in Fig.~\ref{fig.H.eq.Ising}b), 
there is a second peak that, together with the first one, was not previously reported.
In the large temperature limit, $H\to\Hg({\infty})$ and $\Hg({\infty})$ can be 
independently measured on a random configuration with, on average, half of the spins up 
and the other half down. They are shown
as horizontal lines whose height slowly increase with $L$, 
$\Hg({\infty})\sim \log L$~\cite{NoLePa11,JoYiBaKi12}, and are approached as
$\Hg(T)-\Hg({\infty})\sim T^{-1}$. This excess heterogeneity at large
temperatures for the geometric domains is due to the entropic contribution of 
uncorrelated spins. Because of this size dependence, that persists even at very high temperatures
where $\xi\to 0$, the scaling of this second peak turns to be rather complicated. Indeed,
it does not collapse well with exponents obtained for the first peak and, instead, a 
satisfactory collapse was obtained with $\tau\simeq 2.048$ and $\nu\simeq 1.25$. 
These values are close to the random site percolation exponents, 
$\taup=187/91$ and $\nu=4/3$ (thus, $\nuH=187/72$), but this may
be just a crossover because of the large separation between the two peaks for
finite systems (see the 3d case below). 
It seems that even larger systems would be necessary to clarify the
nature of the critical exponents in this problem.
Both peaks, besides increasing in height, move towards $\Tc$ in the thermodynamic
limit. Indeed, both horizontal scalings in Figs.~\ref{fig.H.eq.Ising}d and \ref{fig.H.eq.Ising}e
considered the known value of $\Tc$ in $d=2$. This is a consequence of the fact
that, in $d=2$, the geometric clusters percolative and the thermal transitions occur at 
the same temperature. 

\begin{figure}[htb]
\includegraphics[width=8cm]{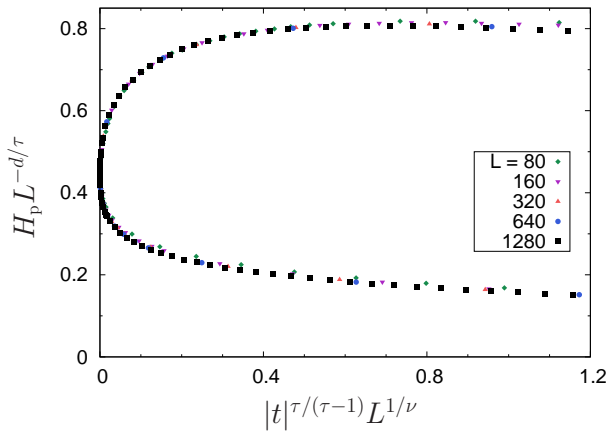}
\caption{Collapsed $\HFK$ for $d=2$, analogous to Fig.~\ref{fig.H.eq.Ising}f, but
using the alternative scaling with $\nu=1$.}
\label{fig.H.eq.Ising.col.PMCO}
\end{figure}

Since $H$ obviously depends on the nature of the clusters, further insight is gained by 
comparing with the behavior of physical clusters. If we remove from the geometric clusters 
the fraction of spins that, albeit parallel, are not effectively correlated, we obtain the 
so-called Coniglio-Klein droplets~\cite{CoKl80,CoFi09} whose physical properties are also
related with the random cluster model~\cite{FoKa72}.
Fig.~\ref{fig.H.eq.Ising}c shows $\HFK$ as a function of the temperature for several sizes
and presents a clear peak close to the transition
that gets larger as the size of the system increases. Notice that the temperatures of all the
three peaks discussed here are well separated for the system sizes considered. 
Differently from $\Hg(\infty)$, that is an increasing function of $L$, 
$\HFK\to 1$ when $T\to\infty$ since all spins become uncorrelated and form
 clusters of unit size.  The scaling proposed in  Ref.~\cite{LeKiPa11} works very well 
also for $\HFK$, Fig.~\ref{fig.H.eq.Ising}f, with the Ising thermal exponents ($\nu=1$,
$\tauFK=31/15$ and $\nuH=31/16$). There is, however, another possible
point of view for the scaling in this problem. Obviously, if $|t|L^{1/\nuH}$ is a 
scaling variable, so is $(|t|L^{1/\nuH})^{\tau/(\tau-1)}$. Indeed, using 
$|t|^{\tau/(\tau-1)}L^{1/\nu}$, collapses equivalent to those in Fig.~\ref{fig.H.eq.Ising} 
are obtained (an example is shown in Fig.~\ref{fig.H.eq.Ising.col.PMCO}), with no need to
resort to a new exponent $\nu_H$. 
In order to collapse the data for $H$, besides understanding how its
peak scales with $L$, one must also know how the width of the critical
region depends on the system size. Standard finite size scaling considers
the region around $\Tc$ in which the measured correlation length $\xi$ will differ from 
its infinite size limit behavior. In a finite system, the divergence of $\xi$ as $T\to\Tc$, 
$\xi\sim |T-\Tc|^{-\nu}$, is constrained by the linear size $L$, and the width of the critical 
region is thus $|T-\Tc| \sim L^{-1/\nu}$.  This interval is, in principle, different for 
other divergent quantities at $\Tc$.  If we consider $H$ instead of $\xi$, since 
$1/\nu=d\sigma/(\tau-1)>d\sigma/\tau$, the divergence of $H$, $H\sim L^{d/\tau}$, is weaker 
and the critical region, wider. Indeed, as discussed in section~\ref{section.H},
$|T-\Tc|\sim L^{-(\tau-1)/\tau\nu}$. 

\begin{figure}[ht]
\includegraphics[width=8cm]{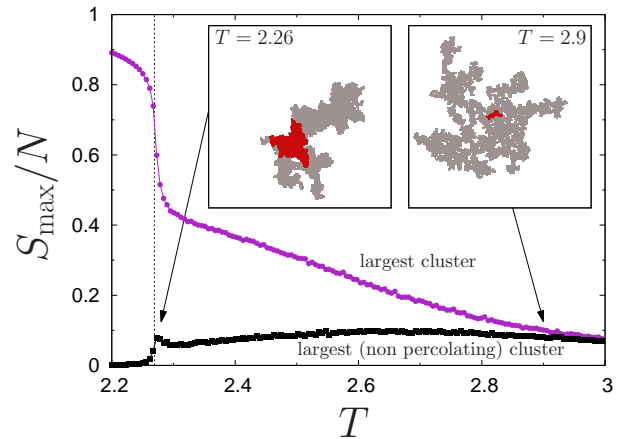}
\caption{Size of the largest (top) and the largest non-percolating (bottom) clusters, normalized
by $N=L^d$, as a 
function of temperature for $L=640$. A cluster is considered percolating when it touches all sides of the
square lattice. Below $\Tc$, indicated by a vertical dashed line, there is almost
always a large, percolating cluster and the available space for the second largest cluster is
small. For large temperatures, it is almost certain that there is no percolating cluster and
both curves coincide. The insets show snapshots for $L=320$ with the largest non-percolating geometric
cluster in gray at two different temperatures (that roughly correspond to the local maxima of
the bottom curve). At the critical temperature, left snapshot, there is a percolating cluster (not
shown) and the second largest is compact. In red we show, among the several physical subclusters in which
it divides, the largest one. The right snapshot shows that, at higher temperature (second maximum), 
although the largest non-percolating geometric cluster has almost the same average size as at
$\Tc$, the physical subclusters are much smaller.}
\label{fig.largest}
\end{figure}

Fig.~\ref{fig.largest} shows the fraction of the whole system occupied by the largest 
cluster as a function of
the temperature, distinguishing whether the largest cluster does or does not percolate. At
very high temperatures both measures coincide since the noise is too high to allow a percolating
cluster. As the temperature decreases, spins become more correlated, the clusters 
larger and these measures, still coinciding, steadily increase. 
They become different when some samples percolate at temperatures roughly below the second peak of
$\Hg$. The average fraction size of the largest component (top
curve) increases almost linearly from this point, as $T$ decreases, up to $\Tc$ where 
there is an abrupt jump and it approaches unity below $\Tc$. It is, on the other hand, 
the other measure, the fraction of the largest non-percolating cluster that seems to
encode the relevant information for $H$ and presents a similar double peak structure.
When some samples start to percolate, the largest non-percolating cluster is actually
the second largest in these samples. As a consequence, its measure starts decreasing,
both measures diverge, leaving a peak close to the temperature of the second peak of $H$. 
The insets of Fig.~\ref{fig.largest} show, in light gray, one example of the largest 
non-percolating cluster at the temperatures indicated, one close to $\Tc$ and the
other at the point of separation of the two measures discussed above. These are 
geometric clusters formed by parallel spins. Each one is a collection of physical
clusters, the largest one is shown in red (dark gray). Obviously, as the temperature
decreases and the spins are more correlated, these clusters increase. 

\begin{figure*}[htb]
\includegraphics[width=8cm]{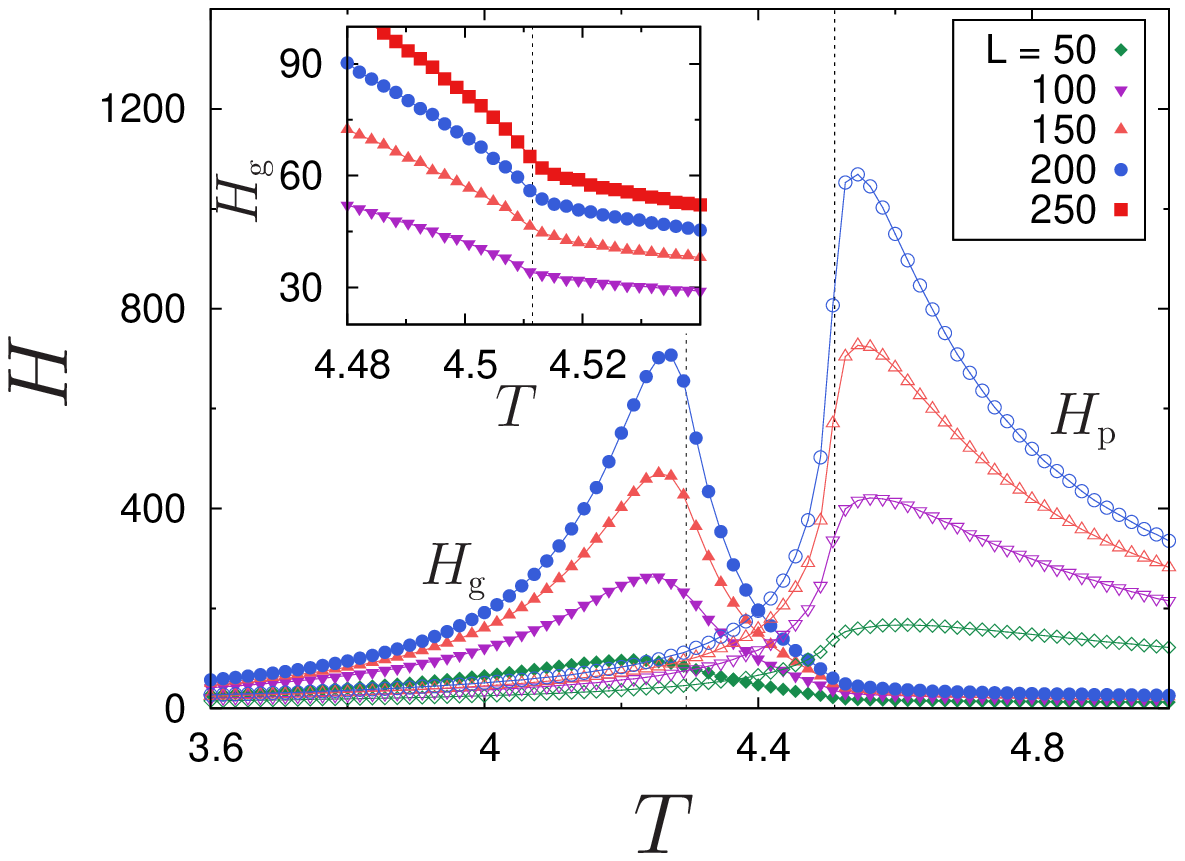}
\includegraphics[width=8cm]{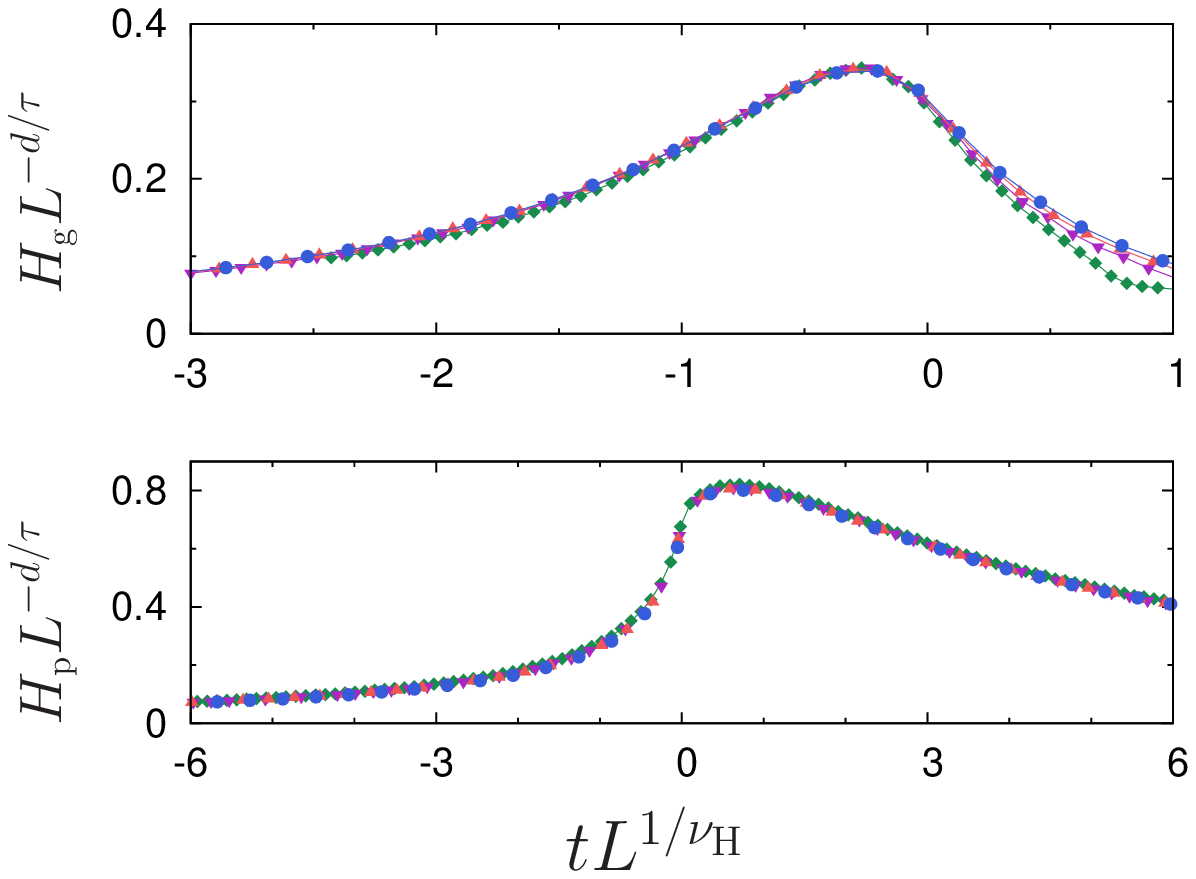}
\caption{(Left) Heterogeneity $H$ for both geometric (left peak, filled symbols) and 
physical clusters (right peak, empty symbols) for the 3d Ising model. The vertical
lines are located at $\Tp\simeq 0.95\Tc$ (pure geometric percolation) and $\Tc\simeq 4.511$ (Ising
critical temperature). Inset: the behavior of $\Hg$ at the thermal critical point, crossing over to a different
declivity. (Right) Collapse of both $\Hg$ (top) and $\HFK$ (bottom). For $\Hg$, $t\equiv (T-\Tp)/\Tp$,
and the best collapse was obtained with $\Tp\simeq 0.95\Tc$, $\tau\simeq 2.08$ and $\nu\simeq 0.75$.
For $\HFK$, on the other hand, $t\equiv (T-\Tc)/\Tc$,
and an excellent collapse is obtained with 3d Ising model exponents: $\nu\simeq 0.875$, $\tau\simeq 2.189$. }
\label{fig.H_3d}
\end{figure*}

\subsection{3d}
\label{section.3d}

The heterogeneities can also be measured in the 3d case, where the
percolative transition of the geometric domains does not coincide
with the thermal transition, $\Tp<\Tc\simeq 4.51$. In 3d, the random site 
percolation threshold for a cubic lattice is 0.312. For all 
temperatures above $\Tc$ (even in the $T\to\infty$ limit), the 
fraction of each spin is 0.5, on average, and since both
spins densities are above the threshold, both percolate. For
temperatures below $\Tc$, the system is magnetized and
the majority spin obviously percolates. 
The percolation transition occurs when, coming from low temperature,
the minority spin (the one opposite to the system magnetization)
whose domains are small and isolate, first percolate. We obtain,
from the best collapse of $H$ (see below),  $\Tp\simeq 0.95\Tc$, 
slightly larger than the estimate in Ref.~\cite{Muller74}, 
$\Tp\simeq 0.92\Tc$. 

Fig.~\ref{fig.H_3d} (left) shows both $\Hg$ and $\HFK$ as a function of temperature. 
Again, $\HFK$ shows a single peak slightly above $\Tc\simeq 4.51$. On the other 
hand, as in the 2d case, $\Hg$ is sensitive to both transitions. While there is 
a large peak very close to $\Tp$, at $\Tc$ there is a crossover to a different
declivity, as shown in the inset of Fig.~\ref{fig.H_3d}. While in the 2d case
two peaks with the same infinite size limit were present, in 3d there is 
a single peak and a crossover, each one located in a different temperature
for $L\to\infty$. The high temperature limit of $\Hg$ seems to increase
even slower than in the 2d case, and is barely visible in the figure.

The data collapses in 3d are shown in Fig.~\ref{fig.H_3d} (right). In the
bottom part, the results for $\HFK$ are very well collapsed, as in the 2d
case, with the thermal exponents of the 3d Ising model and $\Tc\simeq 4.51$: 
$\nu\simeq 0.875$ and $\tau\simeq 2.189$. In the top figure, for $\Hg$,
the best collapse was obtained with $\Tp\simeq 0.95\Tc$, $\tau\simeq 2.08$ 
and $\nu\simeq 0.75$. One would expect~\cite{Fortunato03}, in this specific case, 
the exponents of the 3d random site percolation problem, $\tau_{\scriptscriptstyle\rm perc}
\simeq 2.189$ and $\nu_{\scriptscriptstyle\rm perc}\simeq 0.875$. The 
reason for this~\cite{Fortunato03} is that $\Tp$ being smaller than $\Tc$, the
correlation length is finite at that temperature and clusters at distances
larger than this length are uncorrelated. The small difference between this set
of exponents and those that give the best collapse may be due to finite size
effects~\cite{ArCuPi15}.

\section{Conclusions}
\label{section.conclusions}

The cluster size heterogeneity $H$ is the number of distinct cluster sizes in 
a given sample configuration~\cite{LeKiPa11}. We presented here extensive
simulations for the Ising model for both geometric ($\Hg$) and physical ($\HFK$)
clusters in two and three dimensions. While for the latter, $H$ has a single
peak that diverges at the critical temperature of the Ising model, in the
former case $H$ presents a signature for both the percolative and the thermal
transition. Indeed, finite 2d systems present two quite distinct peaks for $\Hg$ when
the size is sufficiently large, but they merge as $L\to\infty$. On the other
hand, in 3d there is a peak at $\Tp$ and a sudden change in the declivity at $\Tc>\Tp$.
Moreover, since the peaks of $\Hg$ and $\HFK$ do not occur at the same temperature 
in 3d, in the region $\Tp\leq T\leq\Tc$, while $\HFK$ increases, $\Hg$ decreases (a 
similar effect occurs in 2d, but it disappears in the limit $L\to\infty$). 
We also discussed the scaling of $H$ and how its behavior correlates with
the large clusters in the system. Interestingly, although the physical
clusters collapse is clearly related to the Ising thermal exponents, the
behavior of the largest peak of $\Hg$ is more subtle. In both 2d and 3d
the best collapse was obtained with exponents that are close, albeit
different, from those of the respective random site percolation. Although
in 3d this is expected (since $\Tp\neq\Tc$, the correlation length is
finite and distant clusters become uncorrelated~\cite{Fortunato03}), in
2d both transitions coincide and the reason for the similarity with the random site
exponents is still an open question. We conjecture that it may be just
a crossover: for large systems the two peaks get closer to each other
and converge to $\Tc$, the correlation length diverges and even clusters
far apart are correlated. In this way, the exponents should converge
to those of the first peak.

Both in two and three dimensions the heterogeneity measured for geometric
domains, $\Hg$, is very sensitive to the thermodynamical transition, even
if the critical properties are actually encoded in the physical domains. 
The small peak and the declivity change present in 2d and 3d, respectively
are very close to $\Tc$ even for very modest sizes, and its location does 
not suffer from strong finite size effects. If proven general, $\Hg$
may provide a quite precise method to obtain a preliminary estimate for 
the thermal transition.

There are, nonetheless, several remaining questions. For example,
how heterogeneous are the slices of a 3d system~\cite{ArCuPi15}? Differently
from the whole volume behavior, the percolation threshold for the slice
again coincides with $\Tc$. 
Another question is whether the above behavior of $\Hg$ occurs in 
other models as well. We are presently studying the Potts model,
both for continuous and discontinuous transitions, in order to better
understand which are the conditions for having a double peak.
Finally, it is important to perform other geometric measures in
order to better understand the underlying mechanism responsible for
the behavior of $H$. 

\appendix*
\section{On the convergence of Eq.~(\ref{eq.ranking})}
When listing the domain sizes present on a particular configuration,
we notice that all sizes up to a specific size appear, while above
it several holes are present. After averaging over several
configurations, we call this size $s_0$. 

The ranking $r_0$ of the cluster size $s_0$ is
\begin{equation}
r_0 = L^d \sum_{s_0}^\infty n(s)
\end{equation}
not necessarily at the critical point. Near it and using the scaling relation 
$n(s) \simeq s^{-\tau} f(z)$ where $z=ts^\sigma$, one could try to transform this 
discrete sum into an integral $\int_{s_0}^\infty {\rm d}s\,  s^{-\tau} f(z)$. 
Unfortunately, this integral does not converge inside the critical region 
$|t| \sim s_0^{-\sigma} \sim L^{-d\sigma/\tau}$. However, one can 
write the $n$-th derivative of the quoted sum
\begin{eqnarray}
\frac{{\rm d}^n r_0}{{\rm d}t^n} &\simeq& L^d \sum_{s_0}^\infty s^{n\sigma-\tau} f^{(n)}(z)\\
 &=& \frac{L^d}{\sigma}\, t^{-n+\frac{\tau-1}{\sigma}} \int_{z_0}^\infty {\rm d}z\, z^{n-1-\frac{\tau-1}{\sigma}} f^{(n)}(z).\nonumber
\end{eqnarray}
Now, inside the critical region the bottom limit $z_0$ vanishes, and the resulting integral converges to a finite (not $t$-dependent) constant for some high enough integer $n$. 
Therefore, besides additional non-singular terms (proportional to $t^0$, $t^1$ $\dots$ $t^{n-1}$), the leading 
singular form for $r_0$ reads
\begin{equation}
r_0 \sim L^d\, t^{(\tau-1)/\sigma} \sim L^{d/\tau}.
\end{equation}
Notice that the particular mathematical form of the scaling function $f(z)$ does not enter in the argument.
The above scaling relations are valid anyway, only the proportionality prefactors being dependent on $f(z)$. 
In Ref.~\cite{NoLePa11}, the authors assume $f(z)$ to be an exponential function and explicitly use this 
form to derive the above scaling relations. However, in 3 dimensions for instance, $f(z)$ is very well fitted to a (non centered) Gaussian function~\cite{Stauffer86}. In short, the nice and surprising theory introduced in Ref.~\cite{NoLePa11} is correct, but the particular assumption adopted by the authors, in order to justify it, is unnecessary.

\begin{acknowledgments}
JJA deeply thanks Antonio Coniglio for very interesting discussions. 
JJA and PMCO are partially supported by the INCT-Sistemas Complexos and the
Brazilian agencies CNPq and CAPES. JJA has also partial support from FAPERGS. 
\end{acknowledgments}


\begin{thebibliography}{19}
\expandafter\ifx\csname natexlab\endcsname\relax\def\natexlab#1{#1}\fi
\expandafter\ifx\csname bibnamefont\endcsname\relax
  \def\bibnamefont#1{#1}\fi
\expandafter\ifx\csname bibfnamefont\endcsname\relax
  \def\bibfnamefont#1{#1}\fi
\expandafter\ifx\csname citenamefont\endcsname\relax
  \def\citenamefont#1{#1}\fi
\expandafter\ifx\csname url\endcsname\relax
  \def\url#1{\texttt{#1}}\fi
\expandafter\ifx\csname urlprefix\endcsname\relax\def\urlprefix{URL }\fi
\providecommand{\bibinfo}[2]{#2}
\providecommand{\eprint}[2][]{\url{#2}}

\bibitem[{\citenamefont{Cardy and Ziff}(2003)}]{CaZi03}
\bibinfo{author}{\bibfnamefont{J.}~\bibnamefont{Cardy}} \bibnamefont{and}
  \bibinfo{author}{\bibfnamefont{R.~M.} \bibnamefont{Ziff}},
  \bibinfo{journal}{J. Stat. Phys.} \textbf{\bibinfo{volume}{110}},
  \bibinfo{pages}{1} (\bibinfo{year}{2003}).

\bibitem[{\citenamefont{Arenzon et~al.}(2007)\citenamefont{Arenzon, Bray,
  Cugliandolo, and Sicilia}}]{ArBrCuSi07}
\bibinfo{author}{\bibfnamefont{J.~J.} \bibnamefont{Arenzon}},
  \bibinfo{author}{\bibfnamefont{A.~J.} \bibnamefont{Bray}},
  \bibinfo{author}{\bibfnamefont{L.~F.} \bibnamefont{Cugliandolo}},
  \bibnamefont{and} \bibinfo{author}{\bibfnamefont{A.}~\bibnamefont{Sicilia}},
  \bibinfo{journal}{Phys. Rev. Lett.} \textbf{\bibinfo{volume}{98}},
  \bibinfo{pages}{145701} (\bibinfo{year}{2007}).

\bibitem[{\citenamefont{Sicilia et~al.}(2008)\citenamefont{Sicilia, Arenzon,
  Dierking, Bray, Cugliandolo, Martinez-Perdiguero, Alonso, and
  Pintre}}]{SiArDiBrCuMaAlPi08}
\bibinfo{author}{\bibfnamefont{A.}~\bibnamefont{Sicilia}},
  \bibinfo{author}{\bibfnamefont{J.~J.} \bibnamefont{Arenzon}},
  \bibinfo{author}{\bibfnamefont{I.}~\bibnamefont{Dierking}},
  \bibinfo{author}{\bibfnamefont{A.~J.} \bibnamefont{Bray}},
  \bibinfo{author}{\bibfnamefont{L.~F.} \bibnamefont{Cugliandolo}},
  \bibinfo{author}{\bibfnamefont{J.}~\bibnamefont{Martinez-Perdiguero}},
  \bibinfo{author}{\bibfnamefont{I.}~\bibnamefont{Alonso}}, \bibnamefont{and}
  \bibinfo{author}{\bibfnamefont{I.}~\bibnamefont{Pintre}},
  \bibinfo{journal}{Phys. Rev. Lett.} \textbf{\bibinfo{volume}{101}},
  \bibinfo{pages}{197801} (\bibinfo{year}{2008}).

\bibitem[{\citenamefont{Lee et~al.}(2011)\citenamefont{Lee, Kim, and
  Park}}]{LeKiPa11}
\bibinfo{author}{\bibfnamefont{H.~K.} \bibnamefont{Lee}},
  \bibinfo{author}{\bibfnamefont{B.~J.} \bibnamefont{Kim}}, \bibnamefont{and}
  \bibinfo{author}{\bibfnamefont{H.}~\bibnamefont{Park}},
  \bibinfo{journal}{Phys. Rev. E} \textbf{\bibinfo{volume}{84}},
  \bibinfo{pages}{020101} (\bibinfo{year}{2011}).

\bibitem[{\citenamefont{Noh et~al.}(2011)\citenamefont{Noh, Lee, and
  Park}}]{NoLePa11}
\bibinfo{author}{\bibfnamefont{J.~D.} \bibnamefont{Noh}},
  \bibinfo{author}{\bibfnamefont{H.~K.} \bibnamefont{Lee}}, \bibnamefont{and}
  \bibinfo{author}{\bibfnamefont{H.}~\bibnamefont{Park}},
  \bibinfo{journal}{Phys. Rev. E} \textbf{\bibinfo{volume}{84}},
  \bibinfo{pages}{010101} (\bibinfo{year}{2011}).

\bibitem[{\citenamefont{Jo et~al.}(2012)\citenamefont{Jo, Yi, Baek, and
  Kim}}]{JoYiBaKi12}
\bibinfo{author}{\bibfnamefont{W.~S.} \bibnamefont{Jo}},
  \bibinfo{author}{\bibfnamefont{S.~D.} \bibnamefont{Yi}},
  \bibinfo{author}{\bibfnamefont{S.~K.} \bibnamefont{Baek}}, \bibnamefont{and}
  \bibinfo{author}{\bibfnamefont{B.~J.} \bibnamefont{Kim}},
  \bibinfo{journal}{Phys. Rev. E} \textbf{\bibinfo{volume}{86}},
  \bibinfo{pages}{032103} (\bibinfo{year}{2012}).

\bibitem[{\citenamefont{Lv et~al.}(2012)\citenamefont{Lv, Yang, and
  Deng}}]{LvYaDe12}
\bibinfo{author}{\bibfnamefont{J.-P.} \bibnamefont{Lv}},
  \bibinfo{author}{\bibfnamefont{X.}~\bibnamefont{Yang}}, \bibnamefont{and}
  \bibinfo{author}{\bibfnamefont{Y.}~\bibnamefont{Deng}},
  \bibinfo{journal}{Phys. Rev. E} \textbf{\bibinfo{volume}{86}},
  \bibinfo{pages}{022105} (\bibinfo{year}{2012}).

\bibitem[{\citenamefont{Fortuin and Kasteleyn}(1972)}]{FoKa72}
\bibinfo{author}{\bibfnamefont{C.~M.} \bibnamefont{Fortuin}} \bibnamefont{and}
  \bibinfo{author}{\bibfnamefont{P.~W.} \bibnamefont{Kasteleyn}},
  \bibinfo{journal}{Physica} \textbf{\bibinfo{volume}{57}},
  \bibinfo{pages}{536} (\bibinfo{year}{1972}).

\bibitem[{\citenamefont{Coniglio and Klein}(1980)}]{CoKl80}
\bibinfo{author}{\bibfnamefont{A.}~\bibnamefont{Coniglio}} \bibnamefont{and}
  \bibinfo{author}{\bibfnamefont{W.}~\bibnamefont{Klein}}, \bibinfo{journal}{J.
  Phys. A} \textbf{\bibinfo{volume}{13}}, \bibinfo{pages}{2775}
  (\bibinfo{year}{1980}).

\bibitem[{\citenamefont{Newman and Barkema}(1999)}]{NeBa99}
\bibinfo{author}{\bibfnamefont{M.}~\bibnamefont{Newman}} \bibnamefont{and}
  \bibinfo{author}{\bibfnamefont{G.}~\bibnamefont{Barkema}},
  \emph{\bibinfo{title}{Monte Carlo methods in statistical physics}}
  (\bibinfo{publisher}{Oxford University Press}, \bibinfo{address}{New York,
  USA}, \bibinfo{year}{1999}).

\bibitem[{\citenamefont{Stauffer and Aharony}(1994)}]{StAh94}
\bibinfo{author}{\bibfnamefont{D.}~\bibnamefont{Stauffer}} \bibnamefont{and}
  \bibinfo{author}{\bibfnamefont{A.}~\bibnamefont{Aharony}},
  \emph{\bibinfo{title}{Introduction to Percolation Theory}}
  (\bibinfo{publisher}{Taylor \& Francis}, \bibinfo{address}{London},
  \bibinfo{year}{1994}).

\bibitem[{\citenamefont{Stella and Vanderzande}(1989)}]{StVa89}
\bibinfo{author}{\bibfnamefont{A.~L.} \bibnamefont{Stella}} \bibnamefont{and}
  \bibinfo{author}{\bibfnamefont{C.}~\bibnamefont{Vanderzande}},
  \bibinfo{journal}{Phys. Rev. Lett.} \textbf{\bibinfo{volume}{62}},
  \bibinfo{pages}{1067} (\bibinfo{year}{1989}).

\bibitem[{\citenamefont{Janke and Schakel}(2005)}]{JaSc05a}
\bibinfo{author}{\bibfnamefont{W.}~\bibnamefont{Janke}} \bibnamefont{and}
  \bibinfo{author}{\bibfnamefont{A.}~\bibnamefont{Schakel}},
  \bibinfo{journal}{Phys. Rev. E} \textbf{\bibinfo{volume}{71}},
  \bibinfo{pages}{036703} (\bibinfo{year}{2005}).

\bibitem[{\citenamefont{Jan et~al.}(1998)\citenamefont{Jan, Stauffer, and
  Aharony}}]{JaStAh98}
\bibinfo{author}{\bibfnamefont{N.}~\bibnamefont{Jan}},
  \bibinfo{author}{\bibfnamefont{D.}~\bibnamefont{Stauffer}}, \bibnamefont{and}
  \bibinfo{author}{\bibfnamefont{A.}~\bibnamefont{Aharony}},
  \bibinfo{journal}{J. Stat. Phys.} \textbf{\bibinfo{volume}{92}},
  \bibinfo{pages}{325} (\bibinfo{year}{1998}).

\bibitem[{\citenamefont{Coniglio and Fierro}(2009)}]{CoFi09}
\bibinfo{author}{\bibfnamefont{A.}~\bibnamefont{Coniglio}} \bibnamefont{and}
  \bibinfo{author}{\bibfnamefont{A.}~\bibnamefont{Fierro}}, in
  \emph{\bibinfo{booktitle}{Encyclopedia of Complexity and Systems Science}},
  edited by \bibinfo{editor}{\bibfnamefont{R.~A.} \bibnamefont{Meyers}}
  (\bibinfo{publisher}{Springer New York}, \bibinfo{year}{2009}), pp.
  \bibinfo{pages}{1596--1615}.

\bibitem[{\citenamefont{M\"uller-Krumbhaar}(1974)}]{Muller74}
\bibinfo{author}{\bibfnamefont{H.}~\bibnamefont{M\"uller-Krumbhaar}},
  \bibinfo{journal}{Phys. Lett. A} \textbf{\bibinfo{volume}{50}},
  \bibinfo{pages}{27} (\bibinfo{year}{1974}).

\bibitem[{\citenamefont{Fortunato}(2003)}]{Fortunato03}
\bibinfo{author}{\bibfnamefont{S.}~\bibnamefont{Fortunato}},
  \bibinfo{journal}{J. Phys. A: Math.Gen.} \textbf{\bibinfo{volume}{36}},
  \bibinfo{pages}{4269} (\bibinfo{year}{2003}).

\bibitem[{\citenamefont{Arenzon et~al.}(2014)\citenamefont{Arenzon,
  Cugliandolo, and Picco}}]{ArCuPi15}
\bibinfo{author}{\bibfnamefont{J.~J.} \bibnamefont{Arenzon}},
  \bibinfo{author}{\bibfnamefont{L.~F.} \bibnamefont{Cugliandolo}},
  \bibnamefont{and} \bibinfo{author}{\bibfnamefont{M.}~\bibnamefont{Picco}}
  (\bibinfo{year}{2014}), \bibinfo{note}{arXiv/1412.7456}.

\bibitem[{\citenamefont{Stauffer}(1986)}]{Stauffer86}
\bibinfo{author}{\bibfnamefont{D.}~\bibnamefont{Stauffer}}, in
  \emph{\bibinfo{booktitle}{On Growth and Form: Fractal and Non-Fractal
  Patterns in Physics}}, edited by \bibinfo{editor}{\bibfnamefont{H.~E.}
  \bibnamefont{Stanley}} \bibnamefont{and}
  \bibinfo{editor}{\bibfnamefont{N.}~\bibnamefont{Ostrowsky}}
  (\bibinfo{year}{1986}).

\end{thebibliography}

\end{document}